\documentclass{kluwer}   
\usepackage[dvips]{graphicx}

\newcommand{\bm}[1]{\mbox{\boldmath{$#1$}}}
\def\copyrightline{
To appear in {\it {Virtual Astrophysical Jets}}, a special issue of Ap\&SS, \\ 
Kluwer Academic Publishers, 2004.}
\def\idline{}

\newdisplay{guess}{Conjecture}

\begin{document}                                                                                   
\begin{article}
\begin{opening}         
\title
{The efficiency of the magnetic acceleration
\\
in relativistic jets}
\author{Nektarios \surname{Vlahakis}}  
\runningauthor{Nektarios Vlahakis}
\runningtitle{The efficiency of the magnetic acceleration in relativistic jets}
\institute{Section of Astrophysics, Astronomy \& Mechanics, 
Department of Physics, University of Athens, Panepistimiopolis, 
GR-15784 Zografos Athens, Greece}

\begin{abstract}
Using steady, axisymmetric, ideal magnetohydrodynamics (MHD)
we analyze relativistic outflows by means of examining the momentum equation
along the flow and in the transfield direction.
We argue that the asymptotic Lorentz factor is
$\gamma_\infty \sim \mu-\sigma_{\rm M}$,
and the asymptotic value of the 
Poynting-to-matter energy flux ratio -- the so-called $\sigma$ function --
is given by $\sigma_\infty/(1+\sigma_\infty) \sim \sigma_{\rm M} / \mu$,
where $\sigma_{\rm M}$ is the Michel's magnetization parameter
and $\mu c^2$ the total energy-to-mass flux ratio.
We discuss how these values depend on the conditions near the origin of the flow.
By employing self-similar solutions we verify the above result, and show that
a Poynting-dominated flow near the source reaches
equipartition between Poynting and matter energy fluxes,
or even becomes matter-dominated, depending on the value of $\sigma_{\rm M} / \mu$.
\end{abstract}

\end{opening}           

\section{Introduction}
The main driving mechanism for relativistic outflows 
in AGNs, GRBs, and pulsar winds, is likely related to magnetic fields. 
These fields are able to tap the rotational energy of the disk,
and accelerate matter ejecta not only magnetocentrifugally, but also
due to the magnetic pressure.
Using ideal magnetohydrodynamics (MHD) we examine
initially Poynting-dominated outflows, trying to answer the following 
basic question: Which part of the total ejected energy flux
is transfered to the matter kinetic energy flux asymptotically, and how this
value depends on the conditions near the origin of the flow?

The system of equations of special relativistic, steady, cold, ideal MHD,
consist of the Maxwell equations
$0=\nabla \cdot {\bm{B}}
=
\nabla \times {\bm{E}}
=
\nabla \times {\bm{B}} - 4 \pi {\bm{J}} / c
=
\nabla \cdot {\bm{E}} - 4 \pi J^0 / c$,
the Ohm's law 
${\bm{E}}= {\bm{B}} \times {\bm{V}} / c$, 
the continuity 
$\nabla \cdot \left(\rho_0 \gamma {\bm{V}} \right)=0$, 
and momentum 
$-\gamma \rho_0 \left({\bm{V}}\cdot \nabla \right) 
\left(\gamma {\bm{V}}\right)
-\nabla P + \left(J^0 {\bm{E}} +{\bm{J}} \times {\bm{B}} \right) / c =0$
equations.
Here $\bm V$ is the velocity of the outflow,
$\gamma$ the associated Lorentz factor,
$({\bm E}\,,{\bm B})$ 
the electromagnetic field as measured in the central object's frame,
$J^0/c \,,{\bm J}$ the charge and current densities,
and $\rho_0$ the gas rest-mass density in the comoving frame.

Assuming axisymmetry [$\partial/\partial \phi=0$, in 
cylindrical $\left(z\,,\varpi\,,\phi\right)$ coordinates
with $\hat{z}$ along the rotation axis],
four conserved quantities along the flow exist.
If $A=(1/2 \pi) \iint {\bm {B}}_p \cdot d {\bm {S}} $ 
is the poloidal magnetic flux function,
they are
(e.g., \opencite{VK03a})\footnote{
The subscripts $p$/$\phi$ denote poloidal/azimuthal components.}:
the mass-to-magnetic flux ratio
$\Psi_A(A) ={4 \pi \gamma \rho_0 V_p}/{B_p}$,
the field angular velocity
$\Omega(A)= ({V_\phi}/{\varpi})-({V_p}/{\varpi}) ({B_\phi}/{B_p})$,
the specific angular momentum
$L(A)=\gamma \varpi V_\phi -{\varpi B_\phi}/{\Psi_A} $,
and the energy-to-mass flux ratio
\begin{equation}\label{mu}
\mu(A) c^2=\gamma c^2 - {\varpi \Omega B_\phi}/{\Psi_A } \,.
\end{equation}
The right-hand side of eq. (\ref{mu}) consists of the
matter energy-to-mass flux ratio $\gamma c^2$, 
and the Poynting-to-mass flux ratio $(\mu - \gamma) c^2$.
The sum of these two parts is a constant of motion, while their
ratio is defined as the Poynting-to-matter energy flux ratio 
$\sigma = (\mu - \gamma)/\gamma$.

All the physical quantities can be written as functions of $(A\,,\sigma)$:
\begin{equation}\label{bsigma}
B_p = \frac{| \nabla A |}{\varpi} \,,
\quad
B_\phi=-\frac{\sigma}{1+\sigma} \frac{\mu \Psi_A c}{x}\,,
\quad
E=xB_p\,,
\end{equation}
\begin{equation}\label{gammasigma}
\gamma = \frac{\mu}{1+\sigma}\,, 
\quad
\rho_0= \frac{\sigma \Psi_A^2 }{4 \pi x^2} \left[ 
1-\frac{x_{\rm A}^2}{x^2}+ \sigma \frac{ 1-x_{\rm A}^2}{x^2}
\right]^{-1}\,, 
\end{equation}
\begin{equation}\label{vsigma}
\gamma \frac{V_p}{c} = \frac{\Omega^2 \varpi | \nabla A | }{\sigma \Psi_A c^3 }
\left[1-\frac{x_{\rm A}^2}{x^2}+ \sigma \frac{ 1-x_{\rm A}^2}{x^2} \right],
\quad
\gamma \frac{V_\phi}{c}=\mu \frac{x_{\rm A}^2 - \frac{\sigma}{1+\sigma}}{x} ,
\end{equation}
where $x=\varpi \Omega/c$ is the cylindrical distance in units 
of the light cylinder's lever arm, and 
$x_{\rm A}=(L \Omega / \mu c^2)^{1/2}$ its value at the Alfv\'en point.

The functions $A(\varpi\,, z)$, $\sigma(\varpi\,, z)$ obey the two 
remaining equations of the system:
The Bernoulli equation\footnote{This equation comes from
the identity $\gamma^2= 1+(\gamma V_p/c)^2+ (\gamma V_\phi/c)^2$.}
which is a quartic for $\sigma$
\begin{eqnarray}
\left[\frac{\mu}{1+\sigma}\right]^2 \! \! = 1
 \!+ \! \left[ 
\frac{\Omega^2 \varpi | \nabla A | }{\sigma \Psi_A c^3 }
\left(1-\frac{x_{\rm A}^2}{x^2}+ \sigma \frac{ 1-x_{\rm A}^2}{x^2} \right)
\right]^2 
\! \! + \!
\left[\mu \frac{x_{\rm A}^2 - \frac{\sigma}{1+\sigma}}{x} \right]^2
\! \! \! ,
\label{bernoullisigma}
\end{eqnarray}
and the transfield force-balance equation
\begin{eqnarray}
\left[1-\frac{x_{\rm A}^2}{x^2}\right] 
\left[1+\frac{1}{\sigma}\right] 
\frac{\varpi \bar L A }{| \nabla A |}
+2 \frac{\hat{\varpi} \cdot \nabla A}{| \nabla A |}
-\frac{\mu \Psi_A c^6}{\Omega^3 \varpi | \nabla A |}\left[\frac{\sigma}{1+\sigma}\right]^2
\! \!
\frac{d}{dA}\frac{\mu \Psi_A}{\Omega}
\nonumber \\
+\frac{\varpi | \nabla A | }{\Omega} \frac{d \Omega}{dA}
-\frac{1}{\sigma} \left[ 1-\frac{x_{\rm A}^2}{x^2}+ \sigma \frac{ 1-x_{\rm A}^2}{x^2} \right]
\frac{\varpi \nabla A \cdot \nabla \ln \frac{| \nabla A|}{\varpi}}{| \nabla A |}
\nonumber \\
+\frac{\sigma \mu^2 c^4 \Psi_A^2 \left[ x_{\rm A}^2 -\frac{\sigma}{1+\sigma}\right]^2
\hat{\varpi} \cdot \nabla A}
{x^4 \Omega^2 | \nabla A |^3 \left[ 1-\frac{x_{\rm A}^2}{x^2}+ \sigma \frac{ 1-x_{\rm A}^2}{x^2} \right]}
- \frac{\mu^2 c^4 \Psi_A^2 
\sigma \varpi \nabla A \cdot \nabla \sigma
}{x^2 \Omega^2 | \nabla A |^3 (1+\sigma)^3}
=0
\,.
\label{transfieldsigma}
\end{eqnarray}
Here the operator 
$\bar L \equiv \nabla^2 - \frac{2}{\varpi}{\hat{\varpi}} \cdot \nabla $
is related to the curvature radius of the poloidal field lines
${\cal R}= | \nabla A | \left(
\bar L A - \nabla A \cdot \nabla
\ln | \nabla A / \varpi | \right)^{-1}$.

In the force-free limit $\sigma=\infty$ it is
 $\mu =\infty$, $\Psi_A=0$, $L=\infty$, $x_{\rm A}=1$,
while $\mu/\sigma$, $\mu \Psi_A$, $L \Psi_A$, $ \sigma (1-x_{\rm A}^2)$ are finite.
In this case, only the first four terms of
eq. (\ref{transfieldsigma}) survive,
resulting in the ``pulsar equation''
\begin{equation}\label{pulsareq}
\left[1-\frac{1}{x^2}\right] 
\frac{\varpi \bar L A }{| \nabla A |}
+2 \frac{\hat{\varpi} \cdot \nabla A}{| \nabla A |}
-\frac{\mu \Psi_A c^6}{\Omega^3 \varpi | \nabla A |}
\frac{d}{dA}\frac{\mu \Psi_A}{\Omega}
+\frac{\varpi | \nabla A | }{\Omega} \frac{d \Omega}{dA}=0
\,. 
\end{equation}

\section{The $\sigma$ function and its asymptotic value}\label{sigmasection}

\subsection{A general analysis}
An important combination of the field line constants is the
``Michel's magnetization parameter''
$\sigma_{\rm M} (A) = A \Omega^2 / \Psi_A c^3$.
In terms of $\sigma_{\rm M}$, and using
eqs. (\ref{mu}) and (\ref{bsigma}), we may write
the exact expression\footnote{
Eq. (\ref{basiceq}) remains the same with thermal effects included.}
\begin{equation}\label{basiceq}
                                        \framebox{$\displaystyle
\frac{\sigma}{1+\sigma} = 
\left( 
\frac{\sigma_{\rm M}}{\mu}
\right)
\left( 
\frac{-B_\phi}{E}
\right)
\left( 
\frac{B_p \varpi^2}{A}
\right)
                                        $}
\end{equation}
The left-hand side represents the Poynting-to-total energy flux, and 
-- using the first of eqs. (\ref{gammasigma}) --
can be rewritten as $(\mu-\gamma)/\mu$.
As long as the flow is Poynting-dominated 
($\gamma \ll \mu$), this ratio is close to unity, and $\sigma \gg 1$.
This continues to be the case in the neighborhood of the classical fast
magnetosound surface, where $\gamma \approx \mu^{1/3}$ (e.g., \opencite{C86}).
As a result, the superfast regime of the flow is the only
place where a transition from high ($\gg 1$) to $\sim 1$ or $\ll 1$
values of $\sigma$ is possible.
In this regime, and for extremely relativistic flows, the term
($-B_\phi/E$) is very close to unity.\footnote{
The requirement that the Lorentz invariant $B^2-E^2>0$, using $E=xB_p$,
gives $B_\phi^2/E^2> 1- 1/x^2$.
In addition, eqs. (\ref{gammasigma})--(\ref{vsigma}) give
$V_\phi/c= x + (V_p/c)(B_\phi/B_p)$, and the condition $V_\phi>0$ implies
$-B_\phi/E< c/V_p$. Thus, $(1- 1/x^2)^{1/2}<-B_\phi/E< c/V_p$.}
Hence, eq. (\ref{basiceq}) gives a simple relation between the
$\sigma$ function and the ratio $B_p \varpi^2/A$:
\begin{equation}\label{basiceq1}
\frac{\sigma}{1+\sigma} \approx
\left( 
\frac{\sigma_{\rm M}}{\mu}
\right)
\left( 
\frac{B_p \varpi^2}{A}
\right)\,.
\end{equation}
Suppose that the value of the function $B_p \varpi^2/A$ near the 
classical fast surface is $(B_p \varpi^2/A)_f$.
Since $\sigma\gg 1$ at this point,
eq. (\ref{basiceq1}) implies that the constant of motion
$\sigma_{\rm M}/\mu \approx 1/ (B_p \varpi^2/A)_f$.

Denoting with $\delta l_\bot$ the distance between two neighboring 
poloidal field lines
$A$ and $A + \delta A$, magnetic flux conservation implies
$B_p \varpi^2=(\varpi / \delta l_\bot) \delta A$.
Thus,
a decreasing $B_p \varpi^2$ -- and hence, from eq. (\ref{basiceq1}),
a decreasing $\sigma$ function --
corresponds to poloidal field lines expanding
in a way such that their distance $\delta l_\bot$ increases faster than $\varpi$.
How fast the field lines expand is determined by the transfield force balance
equation;
thus, eq. (\ref{transfieldsigma}) indirectly determines the flow acceleration.
Since the available solid angle for expansion is finite,
there is a minimum value of the $B_p \varpi^2/A$ function.
The field lines asymptotically have a shape
$z \approx z_0 (A) + \varpi / \tan \vartheta(A)$, where 
$\vartheta(A)$ is their opening angle.
Differentiating the latter equation we get
$B_p \varpi^2/A= \left( A \vartheta' / \sin \vartheta - A z_0' \sin \vartheta
/ \varpi \right)^{-1}$, a decreasing function, reaching a minimum
value $\sin \vartheta / A \vartheta'$ at $\varpi \gg z_0' \sin^2 \vartheta/
\vartheta'$ \cite{V04}.
Since the factor $\sin \vartheta / A \vartheta'$ is $\sim 1$,
the minimum value of the $B_p \varpi^2/A$ function is $\sim 1$,
corresponding to\footnote{
The only exception to this general result is to have asymptotically
$\sigma/(1+\sigma) \ll \sigma_{\rm M}/\mu$ in some finite solid angle regions,
combined with other regions with bunched field lines
[in which $B_p \varpi^2 \gg A$ and $\sigma/(1+\sigma)\gg \sigma_{\rm M}/\mu$].
Note also that the most general asymptotic field line shape
slightly deviates from straight lines, resulting in a logarithmic acceleration
reaching $\sigma$ values smaller than $\sigma_{\rm M}/\mu$ \cite{CLB91,O02,V04}.
However, this acceleration can happen in exponentially large distances
and hence is physically irrelevant.
}
\begin{equation}\label{basiceq3}
\frac{\sigma_\infty }{1+\sigma_\infty }
\approx
\frac{\sigma_{\rm M}}{\mu} \left(\frac{B_p \varpi^2}{A}\right)_\infty
\approx \frac{(B_p \varpi^2)_\infty }{ (B_p \varpi^2)_f }
\sim \frac{\sigma_{\rm M}}{\mu} 
\sim \frac{1}{ (B_p \varpi^2/A)_f } \,.
\end{equation}
Equivalently, the asymptotic Lorentz factor is $\gamma_\infty 
= \mu / (1+\sigma_\infty)
\sim \mu - \sigma_{\rm M}$,
and the asymptotic Poynting-to-mass flux ratio is $\sim \sigma_{\rm M} c^2$.

Another interesting connection with the boundary conditions near the source
can be found, by noting that, as long as $|B_\phi| \approx E= xB_p$,
$B_p \varpi^2/A \approx 2 |I| / A \Omega$, where $|I|=(c/2) \varpi |B_\phi|$ is the 
poloidal current. Thus, $(B_p \varpi^2/A)_f \approx 2 |I|_f / A \Omega$, and since 
$|I|$ remains constant of motion inside the force-free subfast regime,
$(B_p \varpi^2/A)_f \approx 2 |I|_i /A \Omega$, and $\mu / \sigma_{\rm M} \approx
2 |I|_i /A \Omega$.
Hence, eq. (\ref{basiceq3}) implies a
direct connection of the efficiency and the Lorentz factor to the ejection characteristics
\begin{equation}
\frac{\sigma_\infty }{1+\sigma_\infty }
\approx \frac{A \Omega}{2 |I|_i} \left(\frac{B_p \varpi^2}{A}\right)_\infty
\! \!
\sim  \frac{A \Omega}{2 |I|_i} \,,
\quad \gamma_\infty \sim \mu \left(1 - \frac{A \Omega}{2 |I|_i} \right)
\end{equation}

\subsection{The value of $\sigma_\infty$ in known solutions}
Solving the system of equations (\ref{bernoullisigma}) and (\ref{transfieldsigma})
is highly intractable. In the following we review the 
currently known methods to obtain solutions.

\subsubsection{Numerical methods}
Eq. (\ref{bernoullisigma}) is a relatively simple algebraic equation for $\sigma$.
After substituting its solution (in terms of $A$ and its derivatives) in
the transfield force-balance equation (\ref{transfieldsigma}),
we get a second order partial differential equation for the
magnetic flux function $A$. Its solution determines the
field-streamline shape on the poloidal plane.
Due to the fact that this equation is of mixed type, i.e., changes from elliptic to hyperbolic,
it is beyond the capability of existing numerical codes to solve this 
highly nonlinear problem, and no solution has been obtained so far. 

An alternative numerical approach is to solve the time-dependent problem
(hyperbolic in time) and expect to reach a steady-state. 
However, all existing codes fail to simulate relativistic magnetohydrodynamic
flows for more than a few rotational periods.
On top of that, it is not always clear how the issue of the boundary conditions is handled.

A promising combination of the two above methods is followed by 
\inlinecite{B01}, who solves the inner problem using time-dependent evolution
(avoiding the elliptic to hyperbolic transitions),
and the outer problem using steady-state equations.
The code is not yet capable of solving the problem at large distances, though.

\subsubsection{The force-free assumption}
In the force-free limit $\sigma = \infty$, the two equations 
(\ref{bernoullisigma}) and (\ref{transfieldsigma}) decouple.
Thus, one may solve the somewhat simpler, elliptic
equation (\ref{pulsareq}) (e.g., \opencite{CKF99}),
and then solve eq. (\ref{bernoullisigma}) for $\sigma$.

The force-free solutions have several problems, the most important
of which is the following:
Since the back reaction of the matter to the
field is neglected, the drift velocity soon
after the light cylinder becomes larger than the
light speed. Equivalently, there is no frame of reference where
the electric field vanish, and eq. (\ref{bernoullisigma}) has
no real solutions for $\sigma$.

In any case, the force-free assumption brakes down in the
superfast ($\gamma > \mu^{1/3} \Leftrightarrow
\sigma < \mu^{2/3}-1$) regime, where the flow becomes
hyperbolic and the back reaction of the matter
to the field cannot be neglected. 
Since for Poynting-dominated flows
the value of the $\sigma$ function at the classical fast surface
is $\mu^{2/3}-1 \gg 1$, the force-free method cannot be used for
examining the efficiency of the magnetic acceleration.

\subsubsection{The prescribed field line shape assumption}
If one assumes a known magnetic flux distribution, i.e., a known
function $B_p \varpi^2/A=\varpi |\nabla A|/A$, then it is trivial to solve eq. (\ref{bernoullisigma})
for $\sigma$ and find the flow speed\footnote{In the superfast regime, this equation
reduces to the much simpler eq. [\ref{basiceq1}].} 
(e.g., \opencite{T90}; \opencite{FG01}).
Thus, when we use this method, practically we implicitly give the function $\sigma$!
However, these solutions do not satisfy the transfield force-balance
equation (\ref{transfieldsigma}); thus, they are not fully self-consistent. 

\subsubsection{The monopole approximation}
This is a subcase of the prescribed field line case, based on the
assumption that the poloidal magnetic field is quasi-monopolar.
This assumption, with the help of eq. (\ref{basiceq1}), is 
equivalent to the assumption that the $\sigma$ function is constant!
In fact, a tiny acceleration is possible in the subfast regime,
leading to $\gamma_\infty \sim \mu^{1/3}$ and 
$\sigma_\infty \sim \mu^{2/3} \gg 1$ \cite{M69}.
This solution gave the erroneous impression to the community that relativistic
MHD is in general unable to give high acceleration efficiencies.
However, the solution 
corresponds to a special case of boundary conditions, 
and most importantly, it does not satisfy the transfield force-balance equation.

\subsubsection{Other approximate solutions}
Various tries to solve a simplified version of eqs. (\ref{bernoullisigma})
and (\ref{transfieldsigma}), by neglecting some terms
[e.g., \inlinecite{TT03} who neglected the curvature radius term in 
eq. (\ref{transfieldsigma})]
involve a risk, since the system of equations is highly nonlinear
and it could be crucial to keep terms that seem at first negligible, 
especially second order terms.

Also perturbations around a monopolar solution [e.g.,
\inlinecite{BKR98} who perturbed a solution with $\gamma=\infty$
and $V_\phi=0$] are (using eq. [\ref{basiceq1}]) 
equivalent to perturbations of $\sigma$ around a constant high value.

\medskip
\medskip
Concluding, 
{\bf {in order to solve the efficiency problem,
one has to solve simultaneously 
eqs. (\ref{bernoullisigma}) and (\ref{transfieldsigma})}}.

\subsubsection{The $r$ self-similar special relativistic model}
The only known exact solution of eqs. (\ref{bernoullisigma}) and (\ref{transfieldsigma})
is the $r$ self-similar special relativistic model,
found independently by \inlinecite{LCB92} and 
\inlinecite{C94} in the cold limit, and further 
generalized by \inlinecite{VK03a} including thermal effects.
It corresponds to boundary conditions in a conical surface
($\theta=\theta_i$ in spherical coordinates [$r$, $\theta$, $\phi$]) of the form
$B_r={\cal C}_1 r^{F-2}$,
$B_{\phi}=-{\cal C}_2 r^{F-2}$,
$V_r={\cal C}_3 $,
$V_{\theta}=-{\cal C}_4$,
$V_{\phi}={\cal C}_5$,
$\rho_0={\cal C}_6 r^{2(F-2)}$,
$P={\cal C}_7 r^{2(F-2)}$,
with constant ${\cal C}_1$,$\dots$,${\cal C}_7$.
The parameter of the model $F$ controls the initial current distribution 
[$-\varpi B_\phi = {\cal C}_2 \sin \theta_i \ r^{F-1}$
is an increasing or decreasing function of $r$ for $F>$ or
$<1$; see \inlinecite{VK03a} for details].
Despite the assumed form of the boundary conditions,
the assumption that gravity is negligible, and the absence of 
intrinsic scale, $r$ self-similar remain the only 
self-consistent relativistic MHD solutions.

\begin{figure*}
\centerline{\includegraphics[scale=.74]{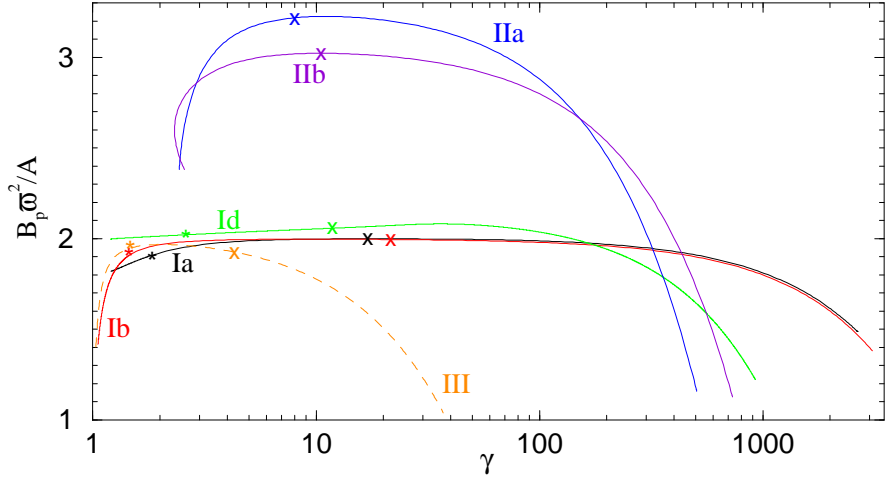}}
\caption{
Function $B_p\varpi^2/A$ {versus} $\gamma$ for exact $r$ self-similar solutions
presented in Vlahakis \& K\"onigl (2003a, 2003b, 2004).
The positions of the Alfv\'en and classical fast magnetosound points
are marked with ``$*$'' and ``$\times$'', respectively.}
\label{figure1}
\end{figure*}
\renewcommand\thefootnote{\alph{footnote}}
\def\arraystretch{.85}
\begin{table*}
\begin{minipage}{1.0\textwidth}
\caption[]{The efficiency in $r$ self-similar models}
\label{table1}
\begin{tabular}{ccccccc}
\hline
 Solution & 
\multicolumn{1}{c}{$\mu$} &
\multicolumn{1}{c}{$\displaystyle \frac{\mu}{\sigma_{\rm M}}$} &
\multicolumn{1}{c}{$\displaystyle \left(\frac{B_p \varpi^2}{A}\right)_f$} &
\multicolumn{1}{c}{$\displaystyle \left(\frac{B_p \varpi^2}{A}\right)_\infty$} &
\multicolumn{1}{c}{$\displaystyle \frac{(B_p \varpi^2)_\infty}{(B_p \varpi^2)_f}$} &
\multicolumn{1}{c}{$\displaystyle \frac{\sigma_\infty }{1+ \sigma_\infty}$} 
\\
\hline
Ia\footnotemark[1] & 10116.1 & 2.02 & 1.99 & 1.49 & 0.74 & 0.73 \\
Ib\footnotemark[1] & 9997.4 & 2.00 & 1.99 & 1.38 & 0.69 & 0.69 \\
Id\footnotemark[1] & 2150.0 & 2.15 & 2.06 & 1.22 & 0.59 & 0.57 \\
IIa\footnotemark[2] & 778.9 & 3.31 & 3.22 & 1.16 & 0.36 & 0.35 \\
IIb\footnotemark[2] & 1156.6 & 3.06 & 3.02 & 1.13 & 0.37 & 0.37 \\
III\footnotemark[3] & 75.0 & 2.06 & 1.93 & 1.04 & 0.54 & 0.51 \\
\end{tabular}
\footnotetext[1]{Trans-Alfv\'enic GRB outflow solutions, from \inlinecite{VK03a}.}
\footnotetext[2]{Super-Alfv\'enic GRB outflow solutions, from \inlinecite{VK03b}.}
\footnotetext[3]{AGN jet solution, from \inlinecite{VK04}.}
\end{minipage}
\end{table*}
\renewcommand\thefootnote{\arabic{footnote}}
\def\arraystretch{1}
Figure \ref{figure1} shows the function $B_p \varpi^2 /A$ for
various $r$ self-similar solutions with application to
GRB and AGN outflows.
With the help of the figure 
and the known $\mu$, $\mu/ \sigma_{\rm M}$ values, 
we are able to fill the table \ref{table1}.\footnote{
Besides the values of $B_p \varpi^2 /A$, fig. \ref{figure1} shows 
$\gamma_\infty$; we also find $\sigma_\infty = (\mu / \gamma_\infty) -1$.}
This table verifies the results of \S~\ref{sigmasection}, and in particular:
1) Since the values of the 3rd and 4th columns are approximately equal,
$\sigma_{\rm M} / \mu \approx 1/(B_p \varpi^2/A)_f$ indeed holds.
2) The value $ (B_p \varpi^2/A)_\infty$ is indeed of order unity.
3) From the two last columns we verify that the asymptotic value of 
$\sigma$ satisfies eq. (\ref{basiceq1}), and 
$\sigma_\infty / (1+ \sigma_\infty) \approx
(B_p \varpi^2/A)_f^{-1}
(B_p \varpi^2/A)_\infty 
\approx (\sigma_{\rm M}/\mu) (B_p \varpi^2/A)_\infty$.
4) The value $\sigma_\infty / (1+ \sigma_\infty) $ is
roughly equal to $\sigma_{\rm M}/\mu$.
5) The value $\gamma_\infty$ is roughly equal to $\mu - \sigma_{\rm M}
=\mu ( 1- \sigma_{\rm M}/\mu)$.
A common characteristic of all trans-Alfv\'enic $r$ self-similar
solutions is that $\mu/ \sigma_{\rm M} \sim 2$, and thus 
$\gamma_\infty \sim \mu /2$.

\subsubsection{$z$ self-similar asymptotic solutions}
\inlinecite{V04} derived $z$ self-similar asymptotic solutions 
of the system of eqs. (\ref{bernoullisigma}) and (\ref{transfieldsigma})
that verify the presented analysis.
For conditions near the classical fast magnetosound surface corresponding
to $(B_p \varpi^2)_f \gg A \Leftrightarrow \sigma_{\rm M} \ll \mu
\Leftrightarrow |I|_i \gg A \Omega /2 
$ the $\sigma_\infty$ is found to be $\ll 1$.

\section{Conclusion}
$\bullet$ In order to solve for the acceleration it is 
absolutely necessary to solve for the poloidal field line shape as 
well. The Bernoulli and transfield force-balance equations are interrelated
and we cannot solve them separately, especially in the superfast regime.

\noindent $\bullet$
Models that assume quasi-monopolar magnetic field
(the Michel's solution included), equivalently
{\emph {assume}} that the magnetic acceleration is inefficient.
Moreover, no fully consistent solution of this kind is known.

\noindent $\bullet$
$\sigma_\infty /(1+\sigma_\infty ) \approx
(\sigma_{\rm M}/\mu) ({B_p \varpi^2}/{A})_\infty
\sim {\sigma_{\rm M}}/{\mu}$ is an analytic expression
for the asymptotic Poynting-to-total energy flux ratio,
confirmed by self-similar solutions. 
Possible acceleration to smaller $\sigma_\infty$
can only happen in exponentially large 
(and thus physically irrelevant) distances.

\noindent $\bullet$
$\sigma_\infty /(1+\sigma_\infty ) \approx
(A \Omega /2 |I|_i) ({B_p \varpi^2}/{A})_\infty
\sim A \Omega /2 |I|_i$ is a general result connecting
the asymptotic value $\sigma_\infty$ (and hence
$\gamma_\infty=\mu/[1+\sigma_\infty]$)
with the conditions near the origin of an
initially Poynting-dominated flow. 

\noindent $\bullet$
The asymptotic Lorentz factor is
$\gamma_\infty \sim \mu - \sigma_{\rm M} \approx \mu (1- A \Omega /2 |I|_i)$,
and the asymptotic Poynting-to-mass flux ratio is $\sim \sigma_{\rm M} c^2
\approx \mu c^2 A \Omega /2 |I|_i$.

\end{article}
\end{document}